\documentclass[useAMS]{mn2e}
\usepackage{epsfig,natbib_vw,times}

\bibpunct[, ]{(}{)}{;}{a}{}{,}

\def \aj {AJ}
\def \mnras {MNRAS}
\def \apj {ApJ}
\def \apjs {ApJS}
\def \apjl {ApJL}
\def \aap {A\&A}

\def \araa {ARAA}

\def \aaps {AAPS}

\def \be {\begin{equation}}
\def \ee {\end{equation}}

\def\gsim{\mathrel{\lower0.6ex\hbox{$\buildrel {\textstyle >}
 \over {\scriptstyle \sim}$}}}
\def\lsim{\mathrel{\lower0.6ex\hbox{$\buildrel {\textstyle <}
 \over {\scriptstyle \sim}$}}}

\def\m@th{\mathsurround=0pt }
\def\eqalign#1{\null\,\vcenter{\openup1\jot \m@th
 \ialign{\strut\hfil$\displaystyle{##}$&$\displaystyle{{}##}$\hfil
 \crcr#1\crcr}}\,}

\def \nca {31~}    
\def \zabs {z_{\rm abs}}
\def \zem {z_{\rm em}}
\def \caii {Ca~{\sc ii}~}
\def \mgii {Mg~{\sc ii}~}
\def \feii {Fe~{\sc ii}~}
\def \mgi {Mg~{\sc i}~}
\def \civ {C~{\sc iv}~}

\title[Dust reddening in DLAs]{Evidence for dust reddening in
  DLAs identified through CaII(H\&K) absorption}
\author[V. Wild \& P. Hewett]{Vivienne Wild\thanks{vw@ast.cam.ac.uk},
  Paul C. Hewett
\vspace*{6pt}\\
Institute of Astronomy, University of Cambridge, Madingley Road,
Cambridge CB3 0HA, UK \\}

\begin{document}
\maketitle

\begin{abstract}
  We present a new sample of \nca \caii(H\&K) $\lambda\lambda
  3935,3970$ absorption line systems with $0.84<\zabs<1.3$ discovered
  in the spectra of Sloan Digital Sky Survey (SDSS) Data Release 3
  quasars, together with an analysis of their dust content.  The
  presence of Calcium absorption together with measurements of the
  \mgii $\lambda 2796$, \feii $\lambda 2600$ and \mgi $\lambda 2853$
  lines lead to the conclusion that the majority of our systems are
  Damped Ly-$\alpha$ (DLA) absorbers.  The composite spectrum in the
  rest frame of the absorber shows clear evidence for reddening.
  Large and Small Magellanic Cloud extinction curves provide
  satisfactory fits, with a best-fit E(B-V) of 0.06, while the
  Galactic dust extinction curve provides a poor fit due to the lack
  of a strong 2175\AA \ feature. A trend of increasing dust content
  with equivalent width of \caii is present. Monte Carlo techniques
  demonstrate that the detection of reddening is significant at
  $>99.99\%$ confidence. The discovery of significant amounts of dust
  in a subsample of DLAs has direct implications for studies of the
  metallicity evolution of the universe and the nature of DLAs in
  relation to high redshift galaxies. The gas:dust ratio is discussed.
  Our results suggest that at least $\sim$40\% of the \caii absorption
  systems are excluded from the magnitude-limited SDSS quasar sample
  as a result of the associated extinction, a fraction similar to the
  upper limit deduced at higher redshifts from radio-selected surveys.
\end{abstract}


\section{Introduction}

The quantity and composition of dust in metal absorption line and damped
Ly-$\alpha$ (DLA) systems has been a topic of debate for over fifteen years.
Quantifying the amount of dust in these systems is important not only
for understanding the chemical evolution of galaxies over large
lookback times, but also has implications for the biases involved in
DLA selection in optical quasar surveys. 

Due to the difficulty in defining suitable samples, there are
relatively few reports of evidence for dust in DLAs via the reddening
effect on background quasar spectra.  Pei, Fall \& Bechtold
(1991)\nocite{1991ApJ...378....6P}, following up from the first study
by Fall, Pei \& McMahon (1989)\nocite{1989ApJ...341L...5F}, found the
spectral energy distributions (SEDs) of 20 quasars with DLAs to be
significantly redder than those of 46 quasars without DLAs.  Recently
however, \citet{2004MNRAS.354L..31M} found no evidence for dust
reddening in a much larger, homogeneous sample of DLAs from the Sloan
Digital Sky Survey (SDSS), finding a limit, E(B-V)$_{\rm{SMC}}<$0.02,
that is inconsistent with the results of Pei et al..  Whilst the
optical selection of the SDSS sample could potentially introduce some
bias into this result, Ellison, Hall \& Lira (2005)\nocite{corals_ebv}
find an E(B-V)$_{\rm{SMC}}<$0.05 (3$\sigma$) using a smaller sample of
radio selected quasars with DLAs.  The alternative view, in which DLAs
possess detectable quantities of dust, is supported by
\citet{1998ApJ...494..175C} who found the paths of 4 out of 5 ``red
quasars'' to contain strong HI 21cm absorption, suggesting the
excessive redness of the quasars to be due to the intervening
absorption systems.

The importance of selection effects in quasar samples when searching
for DLAs or metal absorption line systems is clear, a small amount of
dust present in an intervening galaxy could cause enough extinction
for the quasar to fall below the detection limit of optical
magnitude-limited surveys.  Using a complete sample of radio selected
galaxies containing 22 DLAs with $\zabs>1.8$ in 66 $\zem>2.2$ quasars,
\citet[][the CORALS survey]{2001A&A...379..393E} find that optical
surveys could have underestimated the number of DLAs by at most a
factor of two.  By making use of the criteria for identifying
potential DLAs from the strengths of metal absorption lines
\citep{2000ApJS..130....1R}, \citet{2004ApJ...615..118E} extend this
result to absorbers at lower redshift.  Again their results permit up
to a factor of 2.5 underestimate of the incidence of DLAs in optical,
as opposed to radio selected, quasar samples.

Whilst the intrinsic range in the SEDs of the underlying quasar
spectra means estimates of reddening from differences in the average
SEDs of quasar samples rely on statistical arguments, a clear
indication of the presence of dust comes from the identification of
spectroscopic features caused by the dust grains.  The strongest such
feature in the Galaxy is at $\sim$2175\AA, but the feature is weak or
absent in the Large and Small Magellanic Clouds (LMC and SMC).  The
2175\AA \ feature has been detected in the spectrum of a BL Lac object
at the absorption redshift of a known intervening DLA by
\citet{2004ApJ...614..658J}.  Possible detections in SDSS quasar
spectra with strong intervening metal absorption systems have been
presented by \citet{2004ApJ...609..589W}.

A further diagnostic of the presence of dust in DLAs and metal
absorption line systems comes from relative abundances of elements,
such as Cr to Zn, which are depleted by differing amounts onto dust
grains \citep{1996ARA&A..34..279S}.  Results indicate that dust
depletion is far less severe than in the Galactic interstellar medium
today \citep{1994ApJ...426...79P,1997ApJ...478..536P} which, combined
with results from the radio-selected CORALS survey \citep{CJA05}, has
further strengthened the argument that DLAs are relatively dust free
compared with modern galaxies.

In this paper we analyse a sample of \nca \caii(H\&K) absorption line
systems found in the spectra of SDSS Data Release 3 (DR3) quasars \citep{astro-ph/0503679}.
Calcium is found to be generally underabundant in galaxies and is
expected to be heavily depleted onto dust grains
\citep{1996ARA&A..34..279S}.  Without doubt the detection of \caii
indicates a large column density of HI and a very high fraction of our
sample are expected to be DLAs.  This paper therefore presents one of
the largest homogeneous samples of DLAs at intermediate redshifts and
provides a detection method with an expected success rate
substantially higher than methods based on the properties of \mgii and
\feii absorption.

In Section 2 we present our sample of \caii absorbers.  In Section
\ref{sec:remqso} we describe our method for creating appropriate
``composite'' quasar spectra that allow the statistical removal of the
underlying quasar SEDs to isolate any differences due to the presence
of the \caii absorbers.  Results are given in Section
\ref{sec:results}.  Further details of the method will be presented in
an accompanying paper (Wild, Hewett \& Pettini 2005, in preparation)
together with an analysis of Zn and Cr abundances in the composite
spectrum.

\vspace*{-0.5cm}
\section{A sample of \caii absorbers} \label{sec:sample}

An investigation into the properties of \caii(H\&K: $\lambda\lambda
3934.8, 3969.6$) absorbers at redshifts $z$\,$\sim$\,1 is made
possible by the extended wavelength coverage of the SDSS spectra
(3800-9200\AA) combined with the improved sky subtraction of
\citet{skysub}.
The lower redshift limit to the absorber sample was set
by the appearance of the 2175\AA \ feature at $\lambda>$4000\AA \ in
the SDSS spectra, at $z=0.84$, while the upper redshift limit,
$z=1.3$, was set by \caii(H\&K) moving beyond the red
limit of the spectra.  We further restrict the sample of
\caii absorbers by requiring that the 2175 \AA \ feature falls
redward of the Ly-$\alpha$ forest in the individual quasar spectra.

In order to detect weak \caii(H\&K) absorption features in medium 
resolution spectra, the sample of quasars whose spectra are searched is 
restricted to those obeying the following criteria:
\begin{itemize}
\item entry in the DR3 quasar catalogue \citep{astro-ph/0503679}
\item Galactic extinction corrected $i$-band PSF magnitude $<19.0$
\item spectroscopic signal-to-noise ratio (S/N) in the $i$-band $>10$
\item no broad absorption line (BAL) features.
\end{itemize}
The technique to identify BAL quasars will be presented in detail in
the accompanying paper. Briefly, the scheme involves the calculation of the
root-mean-square (rms) deviations around the \civ, \mgii and \feii lines
from a continuum defined using Principal Component Analysis (PCA).
Thirteen per cent of the quasar spectra, which satisfy
the first three criteria above, are flagged as potential BALs. The final 
sample consists of 11\,427 quasars and
the final conclusions of the paper are insensitive to the precise
scheme used to define the subsample of quasars to be searched. The
faint magnitude limit is similar to that used for the selection of 
``low-redshift'' quasar candidates in the SDSS ($i$-band magnitude of
19.1).   

A matched-filter search \citep{1985MNRAS.213..971H} of the quasar
spectra for \caii doublets (in the ratio of 2:1 and 1:1, matching the
properties of unsaturated and saturated lines) above a 5$\sigma$ significance
threshold was performed.  Candidate systems also had to possess absorption line
detections corresponding to \mgii ($\lambda\lambda 2796,2804$).  Visual
inspection of the candidate list eliminated a small number of
spurious detections.  Table \ref{tab:1} lists the properties of the remaining \nca \caii absorber
candidates.


\begin{table*}
  \centering
\vspace*{-0.4cm}
  \caption{\label{tab:1} \small Name and spectroscopic identification
    of each quasar in our sample, together with rest-frame EWs of \caii(H\&K),
    \mgii($\lambda\lambda 2796,2804$), \mgi($\lambda 2853$) and
    \feii($\lambda 2600$). The final column gives the derived reddening 
    of each quasar (bar one, see text). However, the scatter in the reddening is large due
    to the intrinsic variation in the quasar SEDs from object to
    object, causing some ``negative reddening''s to occur. The PSF magnitudes are
    corrected for Galactic extinction.}
\vspace*{-0.1cm}
\begin{minipage}{16cm}
  \begin{tabular}{cccccccccc} \hline\hline
    SDSS ID & mjd,plate,fibre & mag$_{\rm i}$ & $\zem$ & $\zabs$ & EW \caii &
    EW \mgii & EW \mgi & EW \feii & E(B-V)$_{\rm{LMC}}$  \\ \hline
J002133.36+004301.2 & 51900,0390,537 & 17.42 & 1.245 & 0.942 &  0.34, 0.22 &  1.80, 1.66 &  0.55 &  0.99 & -0.022 \\
J010332.40+133234.8 & 51821,0421,049 & 18.24 & 1.660 & 1.049 &  1.07, 0.80 &  3.06, 2.66 &  1.44 &  2.25 &  0.072 \\
J014717.76+125808.4 & 51820,0429,215 & 17.71 & 1.503 & 1.040 &  0.50, 0.17 &  4.28, 4.26 &  1.15 &  3.05 &  0.041 \\
J074804.08+434138.4 & 51885,0434,340 & 18.44 & 1.836 & 0.898 &  0.53, 0.27 &  1.71, 1.19 &  0.30 &  0.69 &  0.025 \\
J080736.00+304745.6 & 52319,0860,601 & 18.53 & 1.255 & 0.969 &  0.56, 0.79 &  2.83, 2.70 &  1.24 &  2.23 &  0.001 \\
J081054.00+352226.4 & 52378,0892,106 & 18.30 & 1.304 & 0.877 &  0.56, 0.31 &  2.17, 2.11 &  0.99 &  1.78 &  0.032 \\
J081930.24+480827.6 & 51885,0440,007 & 17.64 & 1.994 & 0.903 &  0.75, 0.36 &  1.69, 1.57 &  1.03 &  1.36 & *** \\
J083157.84+363552.8 & 52312,0827,001 & 17.91 & 1.160 & 1.127 &  0.73, 0.41 &  2.52, 2.49 &  0.78 &  1.53 & -0.024 \\
J085221.36+563957.6 & 51900,0448,485 & 18.58 & 1.449 & 0.844 &  0.65, 0.49 &  3.32, 3.03 &  1.24 &  2.45 &  0.010 \\
J085556.64+383231.2 & 52669,1198,100 & 17.57 & 2.065 & 0.852 &  0.45, 0.16 &  2.65, 2.50 &  0.71 &  2.07 &  0.042 \\
J093738.16+562837.2 & 51991,0556,456 & 18.49 & 1.798 & 0.980 &  1.23, 0.62\footnote{Multiple absorption line 
  system. All lines are fit with double Gaussians and the quoted EW is the total of the two systems. 
  When fitting \caii, the velocity-separation of the systems is taken to be that determined from the 
  associated \feii($\lambda 2600$) lines.}
& 4.90, 4.34 &  2.35 &  3.21 &  0.300 \\
J095352.80+080104.8 & 52734,1235,465 & 17.40 & 1.720 & 1.024 &  0.48, 0.35 &  0.91, 0.80 &  0.46 &  0.64 &  0.024 \\
J100000.96+514416.8 & 52400,0903,258 & 18.70 & 1.235 & 0.907 &  0.81, 0.58 &  4.47, 3.90 &  1.41 &  2.47 & -0.012 \\
J100145.12+594008.4 & 52282,0770,087 & 17.82 & 1.186 & 0.900 &  0.47, 0.33 &  0.64, 0.58 &  0.50 &  0.41 &  0.057 \\
J103024.24+561832.4 & 52411,0947,179 & 17.81 & 1.288 & 1.001 &  0.68, 0.33 &  1.91, 1.84 &  0.95 &  1.58 &  0.030 \\
J112053.76+623104.8 & 52295,0775,455 & 17.39 & 1.130 & 1.073 &  0.57, 0.44 &  2.02, 1.94 &  0.92 &  1.50 &  0.070 \\
J112932.64+020422.8 & 51992,0512,113 & 17.31 & 1.193 & 0.966 &  0.56, 0.53 &  2.08, 2.03 &  0.71 &  1.64 & -0.004 \\
J113357.60+510845.6 & 52367,0880,288 & 18.28 & 1.576 & 1.030 &  1.25, 0.67 &  2.66, 2.72 &  0.82 &  1.97 &  0.117 \\
J115244.16+571203.6 & 52765,1311,631 & 17.92 & 1.603 & 0.848 &  0.54, 0.35 &  3.35, 3.19 &  1.15 &  2.33 &  0.212 \\
J122144.64-001142.0 & 52000,0288,078 & 18.52 & 1.750 & 0.929 &  0.58, 0.23 &  0.97, 0.82 &  0.72 &  0.74 & -0.012 \\
J124659.76+030307.2 & 52024,0522,531 & 18.81 & 1.178 & 0.939 &  1.07, 0.60 &  2.91, 2.78 &  1.30 &  2.15 &  0.184 \\
J131058.08+010824.0 & 51985,0295,325 & 17.80 & 1.389 & 0.862 &  0.74, 0.49 &  2.18, 2.27 &  1.33 &  1.46 &  0.209 \\
J144104.80+044348.0 & 52026,0587,329 & 18.42 & 1.112 & 1.040 &  0.97, 0.64 &  2.27, 2.38 &  1.05 &  1.93 &  0.149 \\
J145633.12+544832.4 & 52353,0792,242 & 17.94 & 1.518 & 0.879 &  0.47, 0.45 &  4.02, 3.72 &  1.71 &  3.11 &  0.060 \\
J151247.52+573842.0 & 52079,0612,438 & 18.69 & 2.135 & 1.045 &  0.98, 0.59 &  2.04, 2.28 &  0.75 &  1.57 &  0.139 \\
J153730.96+335837.2 & 52823,1355,633 & 17.44 & 1.024 & 0.913 &  0.38, 0.50 &  1.82, 1.78 &  0.70 &  1.16 &  0.004 \\
J160932.88+462613.2 & 52354,0813,070 & 18.67 & 2.361 & 0.966 &  0.65, 0.36 &  1.07, 0.92 &  0.65 &  0.60 & -0.034 \\
J172739.12+530227.6 & 51821,0359,042 & 17.97 & 1.442 & 0.945 &  0.62,
    0.53 &  2.71, 2.57 &  0.98 &  2.17 & -0.027 \\
J173600.00+573104.8 & 51818,0358,529 & 18.18 & 1.824 & 0.872 &  0.81, 0.60 &  2.01, 1.80 &  0.87 &  1.55 &  0.066 \\
J224511.28+130903.6 & 52520,0739,030 & 18.56 & 1.546 & 0.861 &  2.03, 0.93\footnote{Multiple absorption line system 
  detected in \mgii doublet, for which the total EW of both systems is
    quoted.}  
&  3.99, 3.67 &  1.77 &  3.39 & -0.004 \\
J233917.76-002942.0 & 51877,0385,229 & 18.26 & 1.344 & 0.967 &  0.74, 0.64 &  2.71, 2.44 &  0.85 &  1.80 &  0.108 \\
    \end{tabular}
\vspace*{-0.4cm}
\end{minipage}
\end{table*}

\vspace*{-0.5cm}
\section{Removing the underlying quasar spectra} \label{sec:remqso}

Traditionally the study of reddening in quasar samples has been a
difficult one due to the intrinsic variation in the shape of quasar
SEDs. Using the large sample of quasars in the SDSS survey we can
however obtain a good estimate of the average quasar spectrum.  We do
this by creating composite spectra in redshift bins of $\Delta z =
0.1$ staggered in redshift by $z=0.05$.  Each spectrum is corrected
for Galactic reddening using 
the quoted extinction in the SDSS photometric catalogue and
the Galactic extinction curve from \citet{1989ApJ...345..245C} and
\citet{1994ApJ...422..158O}.  The
spectra are then moved to the quasar rest frame and normalised by
dividing by the median flux in a common wavelength range (avoiding the
main quasar emission lines) before all spectra are combined using an
arithmetic mean.  The statistical power of the SDSS is clear in the
shear number of quasar spectra available for such an analysis.  Up to
a redshift of $\zem=1.9$ more than 500 quasars contribute to each
composite and only for one quasar in our \caii sample does the
appropriate composite depend on less than 100 quasars.

By construction, the composite spectra take account of any systematic
variation in the quasar SEDs as a function of redshift.  It is also
important to take account of any possible variation with magnitude.  A
number of systematic magnitude- or luminosity-dependent effects are
evident in the SDSS quasar spectra \citep{2004AJ....128.2603Y} and
applying a magnitude-dependent correction also has the advantage of
accounting for any potential small systematic variations of the SDSS
spectrophotometry with object flux. For each quasar with detected
\caii absorption we calculate a correction in slope based on its $i$-band
PSF magnitude. The correction is achieved by selecting from those
quasars in the correct redshift range the 80 closest in magnitude. A
second composite is created from these 80 quasars; dividing this
composite by the control spectrum results in a residual spectrum
representing the magnitude dependence, to which we fit a straight
line. On dividing each of our quasar spectra with \caii absorption by
their relevant control spectrum, we then use this fitted line to
account for potential magnitude dependence. Finally, we combine all \nca
spectra, divided by their control spectra and corrected for any
magnitude dependence, into a single composite we term the ``residual
spectrum''.

\vspace*{-0.5cm}
\section{Properties of the sample}\label{sec:results}

In the following subsections we present the measured E(B-V) for the residual
spectrum along with the reddening in each object; a Monte Carlo analysis of the
significance of the result; the equivalent widths (EW) of important metal lines
in the systems and an estimate of the number of DLAs in our sample.

\subsection{The average reddening}

\begin{figure*}
  \begin{minipage}{\textwidth}
    \includegraphics[scale=0.91]{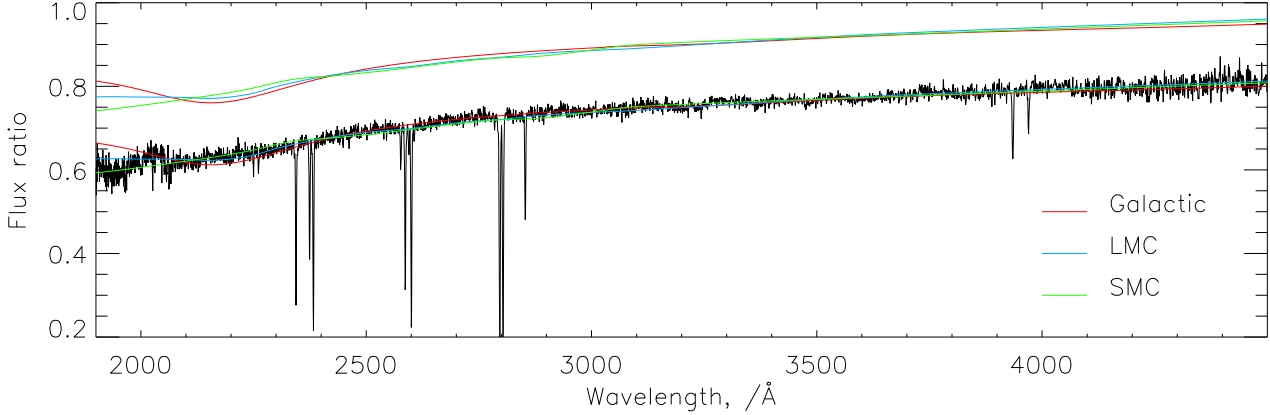}
  \end{minipage}
  \caption{The residual spectrum: a composite of \nca quasars with
    \caii(H\&K) absorption created after division
    of each spectrum by a high S/N quasar control spectrum. Overplotted are
    dust extinction laws for the Galaxy, LMC and SMC. For clarity these
    are also shown offset from the spectrum. The normalisation is
    calculated from that used to fit the LMC dust curve.}
  \label{fig:comp}
\vspace*{-0.2cm}
\end{figure*}

Fig.  \ref{fig:comp} shows the residual spectrum. Each overplotted
dust extinction curve is fitted over a wavelength range of
1900:4500\AA \, excluding the regions containing absorption lines.
The LMC and SMC curves are evaluated from the tabulated results of 
\citet{1992ApJ...395..130P} with an ${\rm R_V}$ of 3.1 and the Galactic
curve is evaluated as in Section \ref{sec:remqso}. The similarity
of the spectrum to the overplotted dust extinction curves is striking.
However, the
Galactic 2175\AA \ dust feature appears to be weak or absent.  The
composite spectrum is consistent with LMC- or SMC-type dust and the
best fit curves have values of E(B-V)=0.056 and 0.057. A jackknife
error of 0.003 is calculated by removing each absorber in turn from the
composite.

\vspace*{-0.2cm}
\subsection{EW \caii(H\&K) vs.  reddening}

\begin{figure*}
\begin{minipage}{\textwidth}
    \includegraphics[scale=0.91]{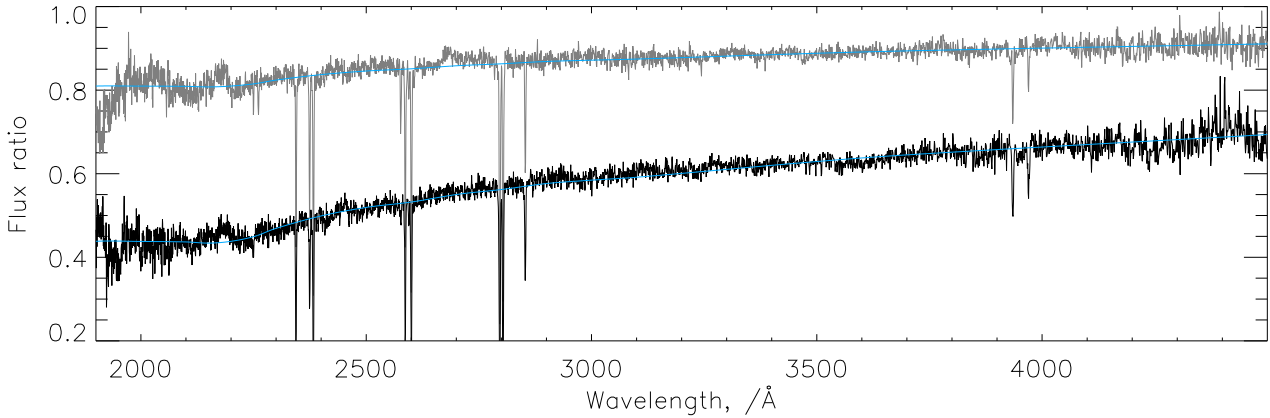}
  \end{minipage}
\caption{Residual spectra created by splitting our sample into
  two: EW(Ca K)$>$0.7\AA \ in black and  EW(Ca K)$<$0.7\AA \ in grey. The large
  (small) EW composite contains 13 (18) 
  objects. Overplotted and offset are the best fitting LMC
  extinction curves.} 
\label{fig:compsplit}
\vspace*{-0.2cm}
\end{figure*}

For each absorber we define a continuum around the \caii lines using
the IRAF routine {\small CONTINUUM} to fit a cubic spline to the
rest-frame wavelength regions 3800:3930, 3940:3965, 3975:4100.  Each
spectrum is fit interactively by altering primarily the order of the
spline until the best by-eye fit is achieved.  The two lines are fit
jointly with Gaussian profiles, fixing the wavelength difference to
the known value and constraining the width of the two lines to be
equal.  Two objects (indicated in the table) show evidence for
multiple absorption systems.  For J093738.16+562837.2 we were able to
fit a double Gaussian to all absorption lines, however, for
J224511.28+130903.6 this was only possible for \mgii.  In all but one
case, in which the quasar is a poor match to the control spectrum, we
can fit a dust curve to each residual spectrum individually.  Although
the results are strongly affected by the variations in the underlying
quasar SEDs, the E(B-V) of the best-fit LMC extinction curve is given
in the final column of Table 1.  Whilst the scatter is large, it is
clear that those spectra with large \caii EWs in general have a large
E(B-V) and are dominating the reddening signal seen in the residual
spectrum (Fig.  \ref{fig:comp}).  Fig.  \ref{fig:compsplit} shows the
two residual spectra created from the sample split by the EW of
\caii(K) at 0.7\AA.  Fitting LMC dust laws to each, we measure
E(B-V)=0.099 and 0.025 for the large and small EW composites
respectively.  The measured E(B-V) values are stable to the precise EW
chosen to split the sample and the conclusions drawn in the final
section of the paper are affected little by moving the boundary.

\vspace*{-0.2cm}
\subsection{Monte Carlo simulations of random samples}

To confirm that our result is not simply due to the variation in
quasar SEDs we create 10~000 Monte Carlo simulations of \nca randomly
chosen quasars, randomly assign them a $\zabs<\zem$ in the same
redshift range as the \caii absorbers, and treat each sample
identically to the actual sample of \caii absorbers.  Fitting an LMC
dust curve to the final residual spectra, we find a maximum E(B-V) of
$<0.04$, i.e. we never obtain an E(B-V) as large as that found in the
Ca II absorber sample and thus obtain a detection significance of
$>99.99\%$. The distribution of simulation E(B-V) has a
variance of $\sim$0.01 and a small tail to redder values, possibly due
to extinction from the host galaxy. 

\vspace*{-0.2cm}
\subsection{Dust obscuration, the proportion of DLAs and HI column density}

Extinction of the quasar due to dust in the intervening absorbers will
cause objects to be missed from the magnitude limited survey.  For
each absorber we calculate the extinction to the background quasar at
7500\AA \ in the observed frame (the effective wavelength of the
$i$-band that defines the magnitude selection) based on
their estimated E(B-V).  The average extinction for all objects is
$\langle A_{7500}\rangle$ = 0.39, with values of 0.67 and 0.18
respectively for the large- and small-EW sub-samples.  Assuming our
ability to detect \caii is unaffected by the magnitude of the
background quasars, we can obtain a lower limit on the fraction of
\caii systems we would have missed due to extinction.  A correction
factor is calculated as the ratio of the total pathlength searched in
quasars within the magnitude limit ($m_{lim}=19$) to the total
pathlength searched in quasars between $m_{lim}$ and $m_{lim}-\langle
A_{7500}\rangle$.  The result is 2.3 for the large-EW sub-sample and
the number of absorbers should therefore be corrected from 13 to 30.
Equivalently, for the small-EW subsample a factor of 1.1 corrects the
number from 18 to 20.  Overall a total corrected number of $\sim50$
\caii systems is therefore expected along the 11\,427 sightlines
searched with an average E(B-V) = 0.07 i.e.  at least 38\% of objects
are missed.  This is consistent with the upper limit expected for DLAs
from the radio selected CORALS survey at high redshift (50\%).  However, our results represent lower limits on the fraction
of systems missed: any objects with a higher dust content will not be
seen in the SDSS quasar sample.

To assess the implications of our result, it is important to
understand the relation between the \caii absorbers and DLAs. Firstly,
converting the minimum EW(Ca K) detected in our sample ($\sim$0.2\AA)
into a column density, N(Ca II) \citep{1948ApJ...108..242S}, we find
Hydrogen column densities of $6 \times 10^{20}$ to $ 2 \times
10^{21}{\rm cm}^{-2}$ are required in our Galaxy to achieve similar
values of N(Ca II) \citep{1993A&AS..100..107S}. The conversion
does not account for the probable lower metallicity of the DLAs and
suggests column densities well in excess of the accepted limit for
DLAs. Secondly, using a sample of 197 absorbers with measured N(HI), Rao,
Turnshek \& Nestor (2005, in preparation) find that, where line EWs
are available, all DLAs have EW(\mgii$\lambda 2796$)/EW(\feii$\lambda
2600$)$<$2 and EW(\mgi$\lambda 2853$)$>$0.2\AA \ and these DLAs
represent 43\% of the \mgii systems fulfilling this criteria.  Also, 8
of 10 systems with EW(2853) $>0.8$\AA \ are confirmed to be DLAs.
Table \ref{tab:1} lists the EWs of metal lines for each of our
absorbers, measured in a similar way to the \caii EWs.  30 of \nca of
our objects fall within the (\mgii,\feii,\mgi) criterion and 20 of \nca
within the \mgi criterion.  These fractions suggest that the majority
of our objects are DLAs according to accepted definitions.

We can also make an estimate of the number density of \caii absorbers
and compare this to values for DLAs.  For our sample of 11\,427
quasars and absorbers with $0.84<\zabs<1.3$ we search a total $\Delta
z \sim 4214$.  Correcting for the effects of dust extinction we
estimate $\sim 50$ absorbers to lie along this path, giving an
$\rm{n(z)} \sim 0.012$.  This is of course only a lower limit as it
does not account for regions in the spectra of poor S/N ratio
prohibiting the identification of \caii.  \citet{2000ApJS..130....1R}
find $\rm{n_{DLA} (z=1.15)} = 0.1^{+0.10}_{-0.08}$.  Performing our
own search for \mgii absorbers in the sample of quasars defined in
Section \ref{sec:sample}, using the \mgii/\feii/\mgi EW criteria
above, we find $\rm{n(z)} = 0.089\pm0.005$.  We conclude that our
results are consistent with strong \caii absorbers representing
$\sim$10\% of all DLAs.

Finally, by assuming an average value of N(HI) for our absorbers
($16.6 \times 10^{20}{\rm cm}^{-2}$, calculated from Table 3 of RT00
for $z_{em} > 0.83$) and taking the obscuration corrected average
E(B-V) of 0.07, we can estimate their gas-to-dust ratio to be $2.4
\times 10^{22}{\rm cm}^{-2}{\rm mag}^{-1}$, which can be compared to
that of the Galaxy; $4.93 \times 10^{21}$ \citep{1994ApJ...427..274D} and
LMC; $2.00 \times 10^{22}$ \citep{1982A&A...107..247K}. If our absorbers were to
contain column densities of neutral Hydrogen in excess of average
DLAs, the value would increase. Multiplying our result by 10
(i.e. assuming the remaining 90\% of DLAs not in our sample contains
negligible dust) we can estimate a global dust-to-gas ratio for DLAs
of $2.4 \times 10^{23}{\rm cm}^{-2}{\rm mag}^{-1}$, in agreement with
the measured lower metallicities of DLAs compared to the Galaxy.

\vspace*{-0.5cm}
\section{Discussion and Conclusions}

Recent results have pointed to DLAs as a sample of chemically young
galaxies with mixed morphologies and evidence for only small
quantities of dust. We have shown that significant reddening due to
dust is present in a subsample of intermediate redshift DLAs
identified through \caii(H\&K) absorption.  The extinction associated
with this detected reddening results in a minimum of $\sim$40\% of
absorbers to be missed from the SDSS magnitude limited sample, a
fraction that is just consistent with the conclusions of the CORALS
radio-selected sample (which relates to the entire DLA population).
The detection of reddening due to dust helps to fill in the gaps in
our understanding of the relation between DLAs and other high redshift
galaxies - our measured E(B-V) values, representing the upper limit
found in DLAs, fit into the lower end of the range covered by Lyman
Break Galaxies at z$\sim$3 \citep[e.g.][]{2001ApJ...562...95S} and
gravitational lens galaxies with $0<z<1$ \citep{1999ApJ...523..617F}.
Follow up observations will add substantially to our knowledge of the
morphology and chemical properties of z$\sim$1 DLAs.

Finally, while criteria based on the properties of \mgii, \feii and
\mgi lines can identify DLAs with a success rate of $\sim$43\%,
selection based on the detection of \caii(H\&K) is likely to increase
the success-rate to $\sim$100\%, albeit at the expense of sensitivity
to only $\sim$10\% of the DLA population.  The two selection
techniques are complimentary and further comparison of samples defined
using both techniques should lead to a greater understanding of the
nature of DLAs.

\vspace*{-0.5cm}
\section*{acknowledgements}
We would like to thank Max Pettini, Michael Murphy, Chris Akerman and
Tae-Sun Kim for valuable discussions and the referee, Joe Liske, for the prompt
and thorough response. Also Sandhya Rao for making
available to us her most recent results on \mgii systems in DLAs and
Chris Akerman and Sara Ellison for their recent results from the
CORALS survey. VW acknowledges the award of a PPARC research
studentship. This work made extensive use of the Craig Markwardt IDL
library.

{\small Funding for the Sloan Digital Sky Survey (SDSS) has been provided by
the Alfred P. Sloan Foundation, the Participating Institutions, the
National Aeronautics and Space Administration, the National Science
Foundation, the U.S. Department of Energy, the Japanese
Monbukagakusho, and the Max Planck Society. The SDSS Web site is
http://www.sdss.org/.

The SDSS is managed by the Astrophysical Research Consortium (ARC) for
the Participating Institutions. The Participating Institutions are The
University of Chicago, Fermilab, the Institute for Advanced Study, the
Japan Participation Group, The Johns Hopkins University, Los Alamos
National Laboratory, the Max-Planck-Institute for Astronomy (MPIA),
the Max-Planck-Institute for Astrophysics (MPA), New Mexico State
University, University of Pittsburgh, Princeton University, the United
States Naval Observatory, and the University of Washington.}

\bibliographystyle{mn2e}

\vspace*{-0.5cm}

\begin{thebibliography}{}
\bibitem[\protect\citeauthoryear{{Akerman}, {Ellison}, {Pettini} \&
  {Steidel}}{{Akerman} et~al.}{2005}]{CJA05}
{Akerman} C.~J.,  {Ellison} S.~L.,  {Pettini} M.,    {Steidel} C.~C.,  2005,
  submitted \aap

\bibitem[\protect\citeauthoryear{{Cardelli}, {Clayton} \& {Mathis}}{{Cardelli}
  et~al.}{1989}]{1989ApJ...345..245C}
{Cardelli} J.~A.,  {Clayton} G.~C.,    {Mathis} J.~S.,  1989, \apj, 345, 245

\bibitem[\protect\citeauthoryear{{Carilli}, {Menten}, {Reid}, {Rupen} \&
  {Yun}}{{Carilli} et~al.}{1998}]{1998ApJ...494..175C}
{Carilli} C.~L.,  {Menten} K.~M.,  {Reid} M.~J.,  {Rupen} M.~P.,    {Yun}
  M.~S.,  1998, \apj, 494, 175

\bibitem[\protect\citeauthoryear{{Diplas} \& {Savage}}{{Diplas} \&
  {Savage}}{1994}]{1994ApJ...427..274D}
{Diplas} A.,  {Savage} B.~D.,  1994, \apj, 427, 274

\bibitem[\protect\citeauthoryear{{Ellison}, {Churchill}, {Rix} \&
  {Pettini}}{{Ellison} et~al.}{2004}]{2004ApJ...615..118E}
{Ellison} S.~L.,  {Churchill} C.~W.,  {Rix} S.~A.,    {Pettini} M.,  2004,
  \apj, 615, 118

\bibitem[\protect\citeauthoryear{{Ellison}, {Hall} \& {Lira}}{{Ellison}
  et~al.}{2005}]{corals_ebv}
{Ellison} S.~L.,  {Hall} P.~B.,    {Lira} P.,  2005, submitted \apj

\bibitem[\protect\citeauthoryear{{Ellison}, {Yan}, {Hook}, {Pettini}, {Wall} \&
  {Shaver}}{{Ellison} et~al.}{2001}]{2001A&A...379..393E}
{Ellison} S.~L.,  {Yan} L.,  {Hook} I.~M.,  {Pettini} M.,  {Wall} J.~V.,
  {Shaver} P.,  2001, \aap, 379, 393

\bibitem[\protect\citeauthoryear{{Falco}, {Impey}, {Kochanek} \& {et
  al.}}{{Falco} et~al.}{1999}]{1999ApJ...523..617F}
{Falco} E.~E.,  {Impey} C.~D.,  {Kochanek} C.~S.,    {et al.} 1999, \apj, 523,
  617

\bibitem[\protect\citeauthoryear{{Fall}, {Pei} \& {McMahon}}{{Fall}
  et~al.}{1989}]{1989ApJ...341L...5F}
{Fall} S.~M.,  {Pei} Y.~C.,    {McMahon} R.~G.,  1989, \apjl, 341, L5

\bibitem[\protect\citeauthoryear{{Hewett}, {Irwin}, {Bunclark}, {Bridgeland},
  {Kibblewhite}, {He} \& {Smith}}{{Hewett} et~al.}{1985}]{1985MNRAS.213..971H}
{Hewett} P.~C.,  {Irwin} M.~J.,  {Bunclark} P.,  {Bridgeland} M.~T.,
  {Kibblewhite} E.~J.,  {He} X.~T.,    {Smith} M.~G.,  1985, \mnras, 213, 971

\bibitem[\protect\citeauthoryear{{Junkkarinen}, {Cohen}, {Beaver}, {Burbidge},
  {Lyons} \& {Madejski}}{{Junkkarinen} et~al.}{2004}]{2004ApJ...614..658J}
{Junkkarinen} V.~T.,  {Cohen} R.~D.,  {Beaver} E.~A.,  {Burbidge} E.~M.,
  {Lyons} R.~W.,    {Madejski} G.,  2004, \apj, 614, 658

\bibitem[\protect\citeauthoryear{{Koornneef}}{{Koornneef}}{1982}]{1982A&A...10%
7..247K}
{Koornneef} J.,  1982, \aap, 107, 247

\bibitem[\protect\citeauthoryear{{Murphy} \& {Liske}}{{Murphy} \&
  {Liske}}{2004}]{2004MNRAS.354L..31M}
{Murphy} M.~T.,  {Liske} J.,  2004, \mnras, 354, L31

\bibitem[\protect\citeauthoryear{{O'Donnell}}{{O'Donnell}}{1994}]{1994ApJ...42%
2..158O}
{O'Donnell} J.~E.,  1994, \apj, 422, 158

\bibitem[\protect\citeauthoryear{{Pei}}{{Pei}}{1992}]{1992ApJ...395..130P}
{Pei} Y.~C.,  1992, \apj, 395, 130

\bibitem[\protect\citeauthoryear{{Pei}, {Fall} \& {Bechtold}}{{Pei}
  et~al.}{1991}]{1991ApJ...378....6P}
{Pei} Y.~C.,  {Fall} S.~M.,    {Bechtold} J.,  1991, \apj, 378, 6

\bibitem[\protect\citeauthoryear{{Pettini}, {King}, {Smith} \&
  {Hunstead}}{{Pettini} et~al.}{1997}]{1997ApJ...478..536P}
{Pettini} M.,  {King} D.~L.,  {Smith} L.~J.,    {Hunstead} R.~W.,  1997, \apj,
  478, 536

\bibitem[\protect\citeauthoryear{{Pettini}, {Smith}, {Hunstead} \&
  {King}}{{Pettini} et~al.}{1994}]{1994ApJ...426...79P}
{Pettini} M.,  {Smith} L.~J.,  {Hunstead} R.~W.,    {King} D.~L.,  1994, \apj,
  426, 79

\bibitem[\protect\citeauthoryear{{Rao} \& {Turnshek}}{{Rao} \&
  {Turnshek}}{2000}]{2000ApJS..130....1R}
{Rao} S.~M.,  {Turnshek} D.~A.,  2000, \apjs, 130, 1

\bibitem[\protect\citeauthoryear{{Savage} \& {Sembach}}{{Savage} \&
  {Sembach}}{1996}]{1996ARA&A..34..279S}
{Savage} B.~D.,  {Sembach} K.~R.,  1996, \araa, 34, 279

\bibitem[\protect\citeauthoryear{{Schneider}, {Hall}, {Richards} \& {et
  al.}}{{Schneider} et~al.}{2005}]{astro-ph/0503679}
{Schneider} D.~P.,  {Hall} P.~B.,  {Richards} G.~T.,    {et al.} 2005,
  astro-ph/0503679

\bibitem[\protect\citeauthoryear{{Sembach}, {Danks} \& {Savage}}{{Sembach}
  et~al.}{1993}]{1993A&AS..100..107S}
{Sembach} K.~R.,  {Danks} A.~C.,    {Savage} B.~D.,  1993, \aaps, 100, 107

\bibitem[\protect\citeauthoryear{{Shapley}, {Steidel}, {Adelberger},
  {Dickinson}, {Giavalisco} \& {Pettini}}{{Shapley}
  et~al.}{2001}]{2001ApJ...562...95S}
{Shapley} A.~E.,  {Steidel} C.~C.,  {Adelberger} K.~L.,  {Dickinson} M.,
  {Giavalisco} M.,    {Pettini} M.,  2001, \apj, 562, 95

\bibitem[\protect\citeauthoryear{{Str{\" o}mgren}}{{Str{\"
  o}mgren}}{1948}]{1948ApJ...108..242S}
{Str{\" o}mgren} B.,  1948, \apj, 108, 242

\bibitem[\protect\citeauthoryear{{Wang}, {Hall}, {Ge}, {Li} \&
  {Schneider}}{{Wang} et~al.}{2004}]{2004ApJ...609..589W}
{Wang} J.,  {Hall} P.~B.,  {Ge} J.,  {Li} A.,    {Schneider} D.~P.,  2004,
  \apj, 609, 589

\bibitem[\protect\citeauthoryear{{Wild} \& {Hewett}}{{Wild} \&
  {Hewett}}{2005}]{skysub}
{Wild} V.,  {Hewett} P.~C.,  2005, \mnras, 358, 1083

\bibitem[\protect\citeauthoryear{{Yip}, {Connolly}, {Vanden Berk} \& {et
  al.}}{{Yip} et~al.}{2004}]{2004AJ....128.2603Y}
{Yip} C.~W.,  {Connolly} A.~J.,  {Vanden Berk} D.~E.,    {et al.} 2004, \aj,
  128, 2603

\end{thebibliography}


\end{document}